\documentclass[runningheads]{llncs} 

\usepackage{float}
\usepackage{graphicx}
\usepackage{wrapfig}
\usepackage{url}
\usepackage[labelfont=bf]{caption}
\makeatletter
\newcommand*{\rom}[1]{\expandafter\@slowromancap\romannumeral #1@}
\makeatother

\usepackage{verbatim}
\usepackage{color}
\usepackage{amsmath} 
\usepackage{mathtools} 
\usepackage{amsfonts} 
\usepackage{amssymb}
\usepackage[normalem]{ulem}

\usepackage{array}

\usepackage{booktabs}
\usepackage{rotating}
\usepackage{multirow}
\usepackage{multicol}
\usepackage{lscape}
\usepackage{subfig}
\usepackage{comment}
\usepackage[export]{adjustbox}

\usepackage{algorithm,algorithmicx,algpseudocode}

\usepackage{lineno}
\usepackage{soul}
\usepackage{color}
\usepackage{xcolor}
\usepackage{xargs}

\definecolor{brilliantrose}{rgb}{1.0, 0.33, 0.64}

\usepackage[colorlinks=true,linkcolor=blue,citecolor=blue]{hyperref}

\usepackage[colorinlistoftodos,prependcaption,textsize=tiny]{todonotes}
\newcommandx{\ISLEM}[2][1=]{\todo[linecolor=brilliantrose,backgroundcolor=brilliantrose!25,bordercolor=brilliantrose,#1]{#2}}

\usepackage{color}

\definecolor{darkgreen}{rgb}{0.53, 0.66, 0.42}

\usepackage{tikz,xcolor,hyperref}

\definecolor{lime}{HTML}{A6CE39}
\DeclareRobustCommand{\orcidicon}{
	\begin{tikzpicture}
	\draw[lime, fill=lime] (0,0) 
	circle [radius=0.16] 
	node[white] {{\fontfamily{qag}\selectfont \tiny ID}};
	\draw[white, fill=white] (-0.0625,0.095) 
	circle [radius=0.007];
	\end{tikzpicture}
	\hspace{-2mm}
}

\foreach \x in {A, ..., Z}{\expandafter\xdef\csname orcid\x\endcsname{\noexpand\href{https://orcid.org/\csname orcidauthor\x\endcsname}
			{\noexpand\orcidicon}}
}
			
\newcommand{\squeezeup}{\vspace{-2.5mm}}

\begin{document}

\title{Inter-Domain Alignment for Predicting High-Resolution Brain Networks Using Teacher-Student Learning}

\titlerunning{Inter-Domain Alignment for High-Resolution Brain Network Synthesis}  

\author{Ba\c{s}ar Demir\orcidC{} \index{Demir, Ba\c{s}ar}\inst{1,\dagger} \and Alaa Bessadok\orcidB{}\index{Bessadok, Alaa}\inst{1,2,3,\dagger}
\and Islem Rekik\orcidA{} \index{Rekik, Islem}\inst{1}\thanks{ {corresponding author: \url{irekik@itu.edu.tr}, \url{http://basira-lab.com}. $\dagger:$ co-first authors.}} }

\institute{$^{1}$ BASIRA Lab, Faculty of Computer and Informatics, Istanbul Technical University, Istanbul, Turkey \\ $^{2}$ Higher Institute of Informatics and Communication Technologies, University of Sousse, Tunisia, 4011 \\ $^{3}$ National Engineering School of Sousse, University of Sousse, LATIS- Laboratory of Advanced Technology and Intelligent Systems, Sousse, Tunisia, 4023}

\authorrunning{B. Demir et al.}  

\maketitle

\squeezeup

\squeezeup
\begin{abstract}
Accurate and automated super-resolution image synthesis is highly desired since it has the great potential to circumvent the need for acquiring high-cost medical scans and a time-consuming preprocessing pipeline of neuroimaging data. However, existing deep learning frameworks are solely designed to predict high-resolution (HR) image from a low-resolution (LR) one, which limits their generalization ability to brain graphs (i.e., connectomes). A small body of works has focused on superresolving brain graphs where the goal is to predict a HR graph from a single LR graph. Although promising, existing works mainly focus on superresolving graphs belonging to the same domain (e.g., functional), overlooking the domain fracture existing between multimodal brain data distributions (e.g., morphological and structural). To this aim, we propose a novel inter-domain adaptation framework namely, Learn to SuperResolve Brain Graphs with Knowledge Distillation Network (L2S-KDnet), which adopts a teacher-student paradigm to superresolve brain graphs. Our teacher network is a graph encoder-decoder that firstly learns the LR brain graph embeddings, and secondly learns how to align the resulting latent representations to the HR ground truth data distribution using an adversarial regularization. Ultimately, it decodes the HR graphs from the aligned embeddings. Next, our student network learns the knowledge of the aligned brain graphs as well as the topological structure of the predicted HR graphs transferred from the teacher. We further leverage the decoder of the teacher to optimize the student network. In such a way, we are not only bringing the learned embeddings from both networks closer to each other but also their predicted HR graphs. L2S-KDnet presents the first TS architecture tailored for brain graph super-resolution synthesis that is based on inter-domain alignment. Our experimental results demonstrate substantial performance gains over benchmark methods. Our code is available at \url{https://github.com/basiralab/L2S-KDnet}.

\end{abstract}

\section{Introduction}

Multi-resolution neuroimaging has spanned several neuroscientific works thanks to the rich and complementary information that it provides \cite{sanchez2018brain,chen2018efficient}. Existing works showed that the diversity in resolution allows early disease diagnosis \cite{mihri,soussia20197}. While super-resolution images provide more details about brain anatomy and function they correspondingly increase the scan time. Consequently, several deep learning-based cross-resolution image synthesis works have been proposed. For instance, \cite{chen2018efficient} combined convolutional neural network with Generative Adversarial Network (GAN) \cite{Goodfellow:2014} to superresolve magnetic resonance imaging (MRI). In addition, \cite{qu2020synthesized} proposed an autoencoder-based architecture for predicting 7T T1-weighted MRI from its 3T counterparts by leveraging both spatial and wavelet domains. While several multi-resolution image synthesis works have been proposed, the road to superresolving brain graphs (i.e., connectomes) is still less traveled \cite{GNNreviewAlaa}. In a brain graph, nodes denote the region of interest (i.e., ROI) and edges denote the connectivity between pairs of ROIs. Thus, each of the LR and HR brain graphs captures unique aspects of the brain to assess the connectivity patterns and functionalities of the brain regions. For instance, cortical brain graphs are generated using time-consuming image processing pipelines that include several steps such as reconstruction, segmentation and parcellation of cortical surfaces \cite{li2013automated}.

To circumvent the need for costly data acquisition and image processing pipelines, a few cross-resolution brain graph synthesis solutions have been proposed \cite{mihri,kubra,megi,gsr}. For instance, \cite{mihri,kubra} predicted HR brain graph of a particular subject given a single LR graph by first selecting its neighboring LR training brain graphs, second performing a weighted averaging for the selected graphs. Specifically, \cite{kubra} relied on manifold learning and correlation analysis techniques to learn how to generate the nearest graph embeddings for a given testing brain graph. With a different perspective, \cite{mihri} estimated a connectional brain template for the LR domain and selected the most similar graphs by computing the residual distance to the estimated template. Although pioneering, these machine learning (ML) based works have two severe limitations: (i) they are designed in a dichotomized manner, with distinct sub-parts of the framework being trained separately. As a result, being agnostic to cumulative errors is obviously important for developing clinically interpretable forecasts, and (ii) they overlook the handling of \emph{domain fracture} problem resulting in the difference in distribution between the LR and HR domains. 

To alleviate these issues, two recent single modality geometric deep learning (GDL) works have been proposed \cite{megi,gsr}, where graph U-net architectures were designed based on a novel graph superresolving propagation rule. In particular, \cite{gsr} firstly learned the embeddings of the LR brain graphs by leveraging graph convolutional network (GCN) \cite{Kipf:2016} and next superresolved the HR graphs using the proposed graph convolution operation. Later on, \cite{megi} was introduced as an extending work of \cite{gsr}, which proposed an adversarial loss to align the predicted HR brain graphs with the ground truth ones. Such framework mainly performed an \emph{intra-domain alignment} as it is learned within the same domain (i.e., HR brain graphs). Besides, it exclusively superresolves unimodal brain graphs where both LR and HR brain graphs are derived from a single modality such as resting state functional MRI (rsfMRI). Therefore, both landmark works could not overcome the second limitation faced in ML-based frameworks either. In this context, several works have demonstrated the impact of domain alignment in boosting the medical image segmentation and reconstruction \cite{zhou2019limited,shen2019brain} which incited researchers in GDL to propose frameworks for aligning graphs \cite{pilavci2019graph,pilanci2020domain}. However, neither the image-based works nor the graph-based ones can be generalized to brain graph super-resolution task. Specifically, to superresolve brain connectomes, one needs to align the LR brain graphs (e.g., morphological) to the HR brain graphs (e.g., functional). However, such an \emph{inter-domain alignment} is strikingly lacking in the super-resolution brain graph synthesis task.

To solve the above challenges, we propose a `Learn to SuperResolve Brain Graphs with Knowledge Distillation Network' (L2S-KDnet) architecture, an affordable artificial intelligence (AI) solution for the brain graph scarcity problem. Essentially, L2S-KDnet is the first inter-domain alignment learning framework designed for brain graph super-resolution rooted in a teacher-student (TS) paradigm \cite{hinton2015distilling}. The goal of a TS framework is to effectively train a simpler network called student, by transferring the knowledge of a pretrained complex network called teacher. It was originally proposed to reduce the computation cost of existing cumbersome deep models encompassing a large number of parameters. Therefore, a TS framework is deployable in real-time applications.

Motivated by such a design, our L2S-KDnet framework consists of two graph encoder-decoder models. The first one is a \emph{domain-aware teacher} which is adversarially regularized using a discriminator, and the second one is a \emph{topology-aware student} network which seeks to preserve the topological properties of the aligned LR to HR graphs and the ground truth graphs. Mainly, we aim to enforce the synthesized HR brain graphs to not only retain the global topological scale of a graph encoded in its connectivity structure but also the local topology of a graph encoded in the hubness of each brain region (i.e., node). Moreover, we propose to leverage the pretrained decoder of the teacher to further boost the learning representation of the student network thereby ensuring that the HR synthesis of the student preserves the topological properties learned by the teacher network. 

\begin{figure}[ht!]
\centering
\hspace{-25pt}
\includegraphics[width=13cm]{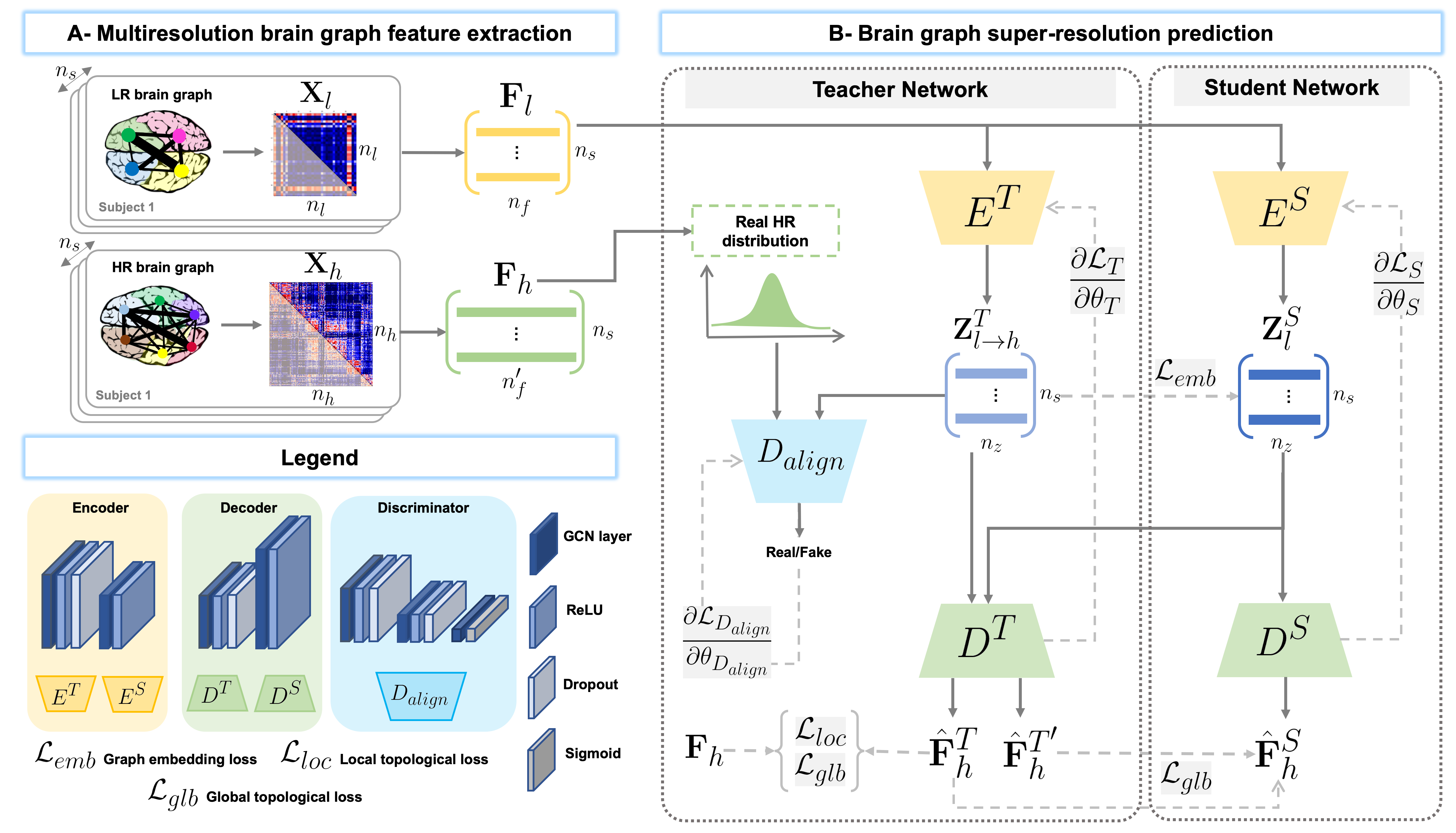}
\caption{ \emph{Proposed pipeline for predicting high-resolution brain graph from a low-resolution one using teacher-student paradigm.} \textbf{(A) Multiresolution brain graph feature extraction.} Each subject is represented by a low-resolution (LR) connectivity matrix $\mathbf{X}_l \in  \mathbb{R}^{n_l \times n_l}$ and a high-resolution (HR) connectivity matrix $\mathbf{X}_h \in  \mathbb{R}^{n_h \times n_h}$ where $n_l$ and $n_h$ represent the resolution of each graph (i.e., number of ROIs) and $n_h > n_l$. Knowing that each brain graph is encoded in a symmetric matrix, we propose to vectorize the upper-triangular part and stack the resulted feature vectors in a matrix $\mathbf{F}_r$ for each brain resolution $r$. \textbf{(B) Brain graph super-resolution prediction.} Firstly, we train a teacher network to align the LR domain to the SR domain in the low-dimensional latent space and we regularize it by a discriminator $D_{align}$. To decode the aligned embeddings, we train the teacher network with a \emph{topology-aware adversarial loss} that so that we enforce the teacher decoder $D^{T}$ to capture both global and local node properties of the real SR graphs. Secondly, we train a student network that aims to accurately predict the SR brain graphs given an LR graph with the knowledge transferred by the teacher. We regularize it with a \emph{topology-aware distillation loss} and we further use the decoder of the teacher to enforce the decoder of the student to give the same prediction results as the teacher network.
}
\label{fig:1}
\end{figure}

\begin{table}
\captionsetup{justification=centering}
\caption{Major mathematical notations used in this paper}
\centering
\begin{scriptsize}
\begin{tabular}{ >{\centering\arraybackslash}m{0.5in} >{\centering\arraybackslash}m{0.5in} >{\centering\arraybackslash}m{3.7in} }
	\toprule
	Notation& 
	Dimension& 
	Definition \\
	\toprule
	$n_s$ & $\mathbb{N}$ & number of training subjects \\
	$n_l$ & $\mathbb{N}$ & number of brain regions (ie. ROIs) in LR brain graph\\
	$n_{h}$ & $\mathbb{N}$ & number of brain regions (ie. ROIs) in HR brain graph\\
	$n_{f}^{\prime}$ & $\mathbb{N}$ &  number of features extracted from the HR brain graph\\
	$n_f$ & $\mathbb{N}$ & number of features extracted from the LR brain graph\\
	$n_{z}$ & $\mathbb{N}$ & number of features of the embedded graphs\\
	$X_l$ & $\mathbb{N}^{n_{l} \times n_{l}}$ & symmetric connectivity matrix representing the LR brain graph\\
	$X_h$ & $\mathbb{N}^{n_{h} \times n_{h}}$ & symmetric connectivity matrix representing the HR brain graph\\
	$Z^{T}_{l\rightarrow h}$ & $\mathbb{N}^{n_{s} \times n_{z}}$ & embedding of the aligned LR to HR brain graphs learned by the teacher's encoder\\
	$Z^{S}_{l}$ & $\mathbb{N}^{n_{s} \times n_{z}}$ & embedding of the LR brain graphs learned by the student's encoder\\
	$F_l$ & $\mathbb{N}^{n_{s} \times n_{f}}$ & feature matrix stacking $n_s$ feature vectors extracted from LR brain graphs \\
	$F_h$ & $\mathbb{N}^{n_{s} \times n_{f}^{\prime}}$ & feature matrix stacking $n_s$ feature vectors extracted from HR brain graphs\\
	$\hat{F}_{h}^{T}$ & $\mathbb{R}^{n_{s} \times n_{f}^{\prime}}$ & HR feature matrix predicted by the teacher network given the teacher's embedding\\
	$\hat{F}_{h}^{T^{\prime}}$ & $\mathbb{R}^{n_{s} \times n_{f}^{\prime}}$ & HR feature matrix predicted by the teacher network given the student's embedding\\
	$\hat{F}_{h}^{S}$ & $\mathbb{R}^{n_{s} \times n_{f}^{\prime}}$ & HR feature matrix predicted by the student network  \\
	$D_{align}$ & $-$ &    discriminator used for inter-domain alignment taking as inputs the real HR distribution extracted from $F_l$ matrix and the embedded LR brain graphs learned by teacher encoder $E^{T}$\\

\bottomrule
\end{tabular}
\end{scriptsize}
\label{tab1}
\end{table}

\section{Methodology}

\textbf{Problem Definition.} A brain graph is a fully connected, weighted and undirected graph that can be represented as \( \mathcal{G}=\{ \mathbf{V},\mathbf{E},\mathbf{X} \}\) where \(\mathbf{V}\) is set of vertices denoting the brain regions, \(\mathbf{E}\) is set of connectivity strengths existing between each pair of vertices and \(\mathbf{X}_{r}\) denotes a connectivity matrix with a specific resolution $r$ and capturing either functional, structural or morphological relationship between two ROIs. Given a LR connectivity matrix $\mathbf{X}_{l} \in  \mathbb{R}^{n_l \times n_l}$ derived from a particular modality (e.g., structural connectivity) we aim to predict a HR $\mathbf{X}_{h} \in  \mathbb{R}^{n_h \times n_h}$ where $n_h > n_l$ that represents another modality (e.g., functional connectivity). We refer the readers to \textbf{Table.}~\ref{tab1} which summarizes the major mathematical notations we used in this paper.

\textbf{A- Multiresolution brain graph feature extraction.} Since each brain graph of resolution $r$ is conventionally encoded in a symmetric connectivity matrix, we propose to reduce the redundancy of edge weights by firstly vectorizing the off-diagonal upper-triangular part using following equation: $n$ = $r \times (r - 1) \over 2$ where $n$ is the dimension of the feature vector and $r$ is the number of nodes (i.e., $r$ can represent $n_l$ or $n_h$). Second, we stack the resulting feature vectors of all subjects into a feature matrix $\mathbf{F}_{r}$. In that way, we get a reduced representation of the connectivity matrices denoting the connectivity strength between pairs of ROIs. Hence, we create two feature matrices representing our graph population with $n_s$ subjects: $\mathbf{F}_{l} \in  \mathbb{R}^{n_s \times n_f}$ and $\mathbf{F}_{h} \in  \mathbb{R}^{n_s \times n_f^{\prime}}$ representing the LR and HR brain graph domains, respectively (\textbf{Fig.}~\ref{fig:1}--A). In other words, we represent the whole population by a single subject-based graph where nodes denote the extracted feature vectors and edges linking pairwise nodes are not weighted. Since our goal is to reduce the redundancy of the weights in the connectivity matrices, we choose to not focus on edges relating nodes thereby defining the adjacency matrix of the graph population as an identity matrix.

\textbf{B- Brain graph super-resolution prediction.} To map the LR brain graph of a subject to the HR graph, we need to first solve the problem of domain fracture resulting in the difference in distribution between both LR and HR domains. Therefore, we propose a TS scheme where we first train a \emph{domain-aware} teacher network aiming to reduce the discrepancy between both domains by learning an \emph{inter-domain alignment} in the low-dimensional space, second, we train a \emph{topology-aware} student network that seeks to preserve the topological properties of both ground truth HR graphs and the aligned brain graph embeddings using the knowledge distilled by the trained teacher. We design our teacher network as a graph GAN composed of a  generator designed as a graph encoder-decoder and a single discriminator. On the other hand, we design our student network as a graph encoder-decoder regularized using three losses which we will explain in detail in the following section. Specifically, each graph encoder-decoder network is composed of an encoder that learns the LR brain graph embeddings, and a decoder that predicts the HR brain graphs. Specifically, we use GCN layers \cite{Kipf:2016} for building all components of our pipeline. 

$\bullet$ \textbf{\emph{Teacher network.}} To align the LR brain graphs to the HR graphs, our teacher network firstly maps the LR graphs into a low-dimensional space using the encoder $E^{T}(\mathbf{F}_{l},\mathbf{A})$. To eliminate redundancy of edge weights, we set the second input to the GCN layer which is the adjacency matrix to an identity matrix $\mathbf{I}$. We build both encoder and decoder with the same sequence of layers, we stack two GCN layers followed by Rectified Linear Unit (ReLU) and dropout function. We define the graph convolution operation proposed by \cite{Kipf:2016} and employed by the GCN layers is defined as follows:

\begin{equation}
\mathbf{Z}^{(l)} = f_{ReLU}(\mathbf{F}^{(l)}, \mathbf{A} \vert \mathbf{W}^{(l)}) = {ReLU}(\mathbf{\widetilde{D}}^{-\frac{1}{2}}\mathbf{\widetilde{\mathbf{A}}}\mathbf{\widetilde{D}}^{-\frac{1}{2}}\mathbf{F}^{(l)}\mathbf{W}^{(l)})
\label{eqn:equation1}
\end{equation}

$\mathbf{Z}^{(l)}$ is the learned graph embedding resulting from the layer $l$ which will be considered as the aligned LR to HR embeddings when training our discriminator. In the first layer of our encoder $E^{T}(\mathbf{F}_{l},\mathbf{I})$ we define $\mathbf{F}^{(l)}$ as the feature matrix of the LR domain $\mathbf{F}_{l}$ while we define it in the second layer as the learned embeddings $\mathbf{Z}$. We denote by $\mathbf{W}^{(l)}$ the learnable parameter for each layer $l$. We define the graph convolution function as $f_{(.)}$ where $\mathbf{\widetilde{\mathbf{A}}} = \mathbf{\mathbf{A}} + \mathbf{I}$ with $\mathbf{I}$ being an identity matrix, and $\mathbf{\widetilde{D}}_{ii} = \sum_{j}\mathbf{\widetilde{\mathbf{A}}_{ij}}$ is a diagonal matrix (\textbf{Fig.}~\ref{fig:1}--B). Since our goal is to enforce the encoder $E^{T}$ to learn an \emph{inter-domain} alignment, we design a discriminator $D_{align}$ to bridge the gap between the LR brain graph embedding distribution and $\mathbf{Z}$ the original HR brain graph distribution. We optimize it using the following loss function:

\begin{equation}
    \max_{D_{align}} \mathbb{E}_{\mathbf{F}^{\prime}_{h} \sim \mathbb{P}_{{\mathbf{F}}_{h}}} [\log D_{align}(\mathbf{F}^{\prime}_{h})] + \mathbb{E}_{\mathbf{Z}_{l\rightarrow h} \sim \mathbb{P}_{\mathbf{Z}}} [\log(1- D_{align}(\mathbf{Z}_{l\rightarrow h}))]
\label{eqn:equation2}
\end{equation}

\noindent where $\mathbb{P}_{{\mathbf{F}}_{h}}$ is the real HR graph distribution, and the distribution $\mathbb{P}_{\mathbf{Z}}$ is the generated distribution by our encoder $E^{T}$ representing the learned aligned LR to HR. On the other hand, our decoder $D^{T}(\mathbf{Z}^{(l)},\mathbf{I})$ which has a reversed structure of the encoder takes the learned embedding as input and superresolves it to get the predicted feature matrix of the HR domain $\hat{\mathbf{F}}_{h}^{T}$. Particularly, both first and second GCN layers play a scaling role since they map the initial dimension of the graph embedding domain into the HR domain. In another word, our decoder can superresolve the brain graphs given the aligned LR brain graph embedding to the HR ground truth data distribution.

Owing to prior works \cite{Bassett:2017,Borgatti:2006} showing the unique topological properties that have a brain connectome such as hubness and modularity, it is important to preserve such properties when superresolving the brain graphs. To achieve this goal, we proposed to regularize the graph encoder-decoder using a novel \emph{topology loss function} $\mathcal{L}_{top}$ that is composed by both local and global topology losses. Specifically, the global topology loss aims to learn the \emph{global graph structure} such as the number of vertices and edges. So, we define it as the mean absolute difference (MAE) between the predicted and ground truth brain features. Moreover, our local topology loss aims to preserve the \emph{local graph structure} which reflects the node importance in the graph. Hence, we define it as the absolute difference in the centrality metrics of nodes. Specifically, we consider the node strength distance between the ground truth and the predicted connectivity matrix $\mathbf{X}_h$ and $\hat{\mathbf{X}}_h$, respectively. We generate the predicted HR matrix by antivectorizing the feature matrix $\hat{\mathbf{F}}_h^T$ generated by the decoder $D^{T}$. Thus, we compute the node strength vector for each subject by summing and normalizing over the rows of the corresponding brain graph. Then, we stack them vertically to obtain $\mathbf{S}_h$ and $\mathbf{\hat{S}}_h$ by summing and normalizing over the rows of $\mathbf{X}_s^{m,r_k}$ and $\mathbf{\hat{X}}_s^{m,r_k}$, respectively. Ultimately, we define the \emph{topology loss function} $\mathcal{L}_{top}$ to train the teacher network as follows: 

\begin{equation}
 	\mathcal{L}_{top} = \sigma \underbrace{\ell_{MAE}(\mathbf{S}_{h}^{T}, \hat{\mathbf{S}}_{h}^{T})}_{\textbf{local topology loss}} + \underbrace{ \ell_{MAE}(\mathbf{F}_{h}^{T}, \hat{\mathbf{F}}_{h}^{T})}_{\textbf{global topology loss}}
\label{eq:topo}
\end{equation}

Note that our graph encoder-decoder acts as a generator in a GAN, thus this boils down to computing the adversarial loss function as follows:

\begin{equation}
    \mathcal{L}_{adv} = \min_{D_{align}}\mathbb{E}_{\mathbf{Z}_{l\rightarrow h} \sim \mathbb{P}_{\mathbf{Z}}} [\log(1- D_{align}(\mathbf{Z}_{l\rightarrow h}))]
\label{eq:adv}
\end{equation}

Given the above definitions of the topology loss \textbf{Eq.}~\ref{eq:topo}, the adversarial \textbf{Eq.}~\ref{eq:adv} loss in addition to a hyper-parameter $\lambda$ that controls the relative importance of the topology loss, we introduce the overall \emph{domain-aware adversarial loss} function of our teacher's generator as follows:

\begin{equation}
 	\mathcal{L}_{T} =  \mathcal{L}_{adv} + \lambda \mathcal{L}_{top}
\label{eq:lossT}
\end{equation}

In that way, we unprecedentedly enforce our teacher network to superresolve brain graphs while jointly solving two critical problems \cite{GNNreviewAlaa}: the \emph{inter-domain} alignment and the \emph{topological property preservation} of brain graphs (\textbf{Fig.}~\ref{fig:1}--B).

$\bullet$ \textbf{\emph{Student network.}} Recall that the ultimate goal of this work is to predict a HR brain graph from a LR graph each derived from a specific modality (e.g., functional and structural). However, there exists an obvious domain gap between the LR and HR domains, which raises the question of whether we can have a model performing an \emph{inter-domain alignment} in that way we ensure that the superresolved graph distribution matches the distribution of the ground truth HR graph. To approach this, we propose a TS learning-based framework where the student network takes advantage of a pretrained \emph{domain-aware} teacher network's knowledge. In that way, our efficient and effective lightweight student model can be easily deployed in real-world clinical applications \cite{hinton2015distilling}. Therefore, we propose to freeze the teacher network to distill its knowledge that will be used to train the student model. Knowing that our teacher is trained using the topology loss function, we assume that the knowledge transferred to the student already encompasses the information related to the local topological properties of the HR brain graphs. Thus, we propose to solely train the student to preserve the global structure of the teacher's predicted graphs $\hat{\mathbf{F}}_{h}^{T}$. Additionally, since the teacher and the student take the same input data that is the LR feature matrix $\mathbf{F}_{l}$ and perform a graph embedding using their GCN encoders, we make the assumption that their latent space should be similar. To ensure that, we further train the student network with additional information of the teacher. More specifically, we propose to regularize it with two global topology-based losses $\mathcal{L}_{emb}$ and $\mathcal{L}_{glb}$ related to the aligned brain graph embeddings and the predicted HR brain graphs, respectively. We define both loss functions optimizing the student network as follows:

\begin{equation}
 	\mathcal{L}_{emb} = \ell_{MAE}(\mathbf{Z}_{l\rightarrow h}^{T}, \mathbf{Z}_{l}^{S});
    \quad\text{ }
    \mathcal{L}_{glb} = \ell_{MAE}(\hat{\mathbf{F}}_{h}^{S}, \hat{\mathbf{F}}_{h}^{T})
\label{eq:10}
\end{equation}

\noindent where $\mathbf{Z}_{l\rightarrow h}^{T}$ and $\mathbf{Z}_{l}^{S}$ denotes the aligned LR to HR graph embeddings learned by the teacher's encoder $E^{T}$ and the learned LR graph embeddings learned by the student's encoder $E^{S}$. $\hat{\mathbf{F}}_{h}^{S}$ and $\hat{\mathbf{F}}_{h}^{T}$ are the predicted HR brain feature matrices of the student and teacher, respectively. By doing so, we enforce the student to preserve both global and local topological properties of not only the synthesized HR graphs from the teacher but also those of the aligned brain graphs. 

Although we want the embedding of the teacher and student to be similar as well as their predicted graphs, still their decoders are not the same. There is no communication between the prediction result between the teacher and student. Therefore, training our student network only with the two losses defined in \textbf{Eq.}~\ref{eq:10} means that we are bringing the lower representations of both teacher and student networks closer to each other independently of the decoder outputs of both networks. However, our goal is to obtain the same HR graph prediction from both networks during the testing. Therefore, inspired by a recent work \cite{yang2021knowledge}, we propose to bridge the gap between the teacher and the student embeddings in such a way this increases the HR brain graph prediction result of the student. Hence, we propose to leverage the pretrained decoder $D^{T}$ of the trained teacher network to further enhance the student loss function. We pass the learned graph embeddings of the student $\mathbf{Z}_{l}^{S}$ to the decoder of the teacher $D^{T}$ which will generate the HR predicted graphs  $\hat{\mathbf{F}}_{\mathrm{h}}^{\mathrm{T}^{\prime}}$. Next, we compute an additional global topology loss between the prediction of the student and teacher decoders and we express our full \emph{topology-aware distillation loss} function as follows:

\begin{equation}
 	\mathcal{L}_{S} =  mean[\lambda_1 \ell_{MAE}(\hat{\mathbf{F}}_{h}^{S}, \hat{\mathbf{F}}_{\mathrm{h}}^{\mathrm{T}^{\prime}}) + \lambda_2 \mathcal{L}_{glb} + \lambda_3 \mathcal{L}_{emb}]
\label{eq:lossS}
\end{equation}

\section{Results and Discussion}

\textbf{Connectomic dataset and parameter setting.} We evaluated our framework on a connectomic dataset derived from the Southwest University Longitudinal Imaging Multimodal (SLIM) Brain Data Repository \cite{qiusouthwest}. It consists of 276 subjects each represented by a morphological LR brain graph of size $35 \times 35$ derived from T1-weighted MRIs and a functional HR brain graph of size $160 \times 160$ derived from resting-state functional MRI. Our encoder comprises two hidden layers that contains 100 and 50 neurons, respectively. We construct discriminator with 3 hidden layers and each contains 32, 16, 1 neurons. We train our model with 150 iterations using 0.0001 as learning rate and  $\beta_{1} = 0.5$, $\beta_{2} = 0.999$ as Adam optimizer parameters for both networks. We used grid search to set our hyperparameters empirically of $\mathcal{L}_{T}$ and $\mathcal{L}_{S}$ and we set $\sigma$ to 0.1, $\lambda$ to 0.5, $\lambda_1$, $\lambda_2$ and $\lambda_3$ are all set to $1$.

\textbf{Comparison methods and evaluation.} Using a single Tesla V100 GPU (NVIDIA GeForce GTX TITAN with 32GB memory), we evaluate our framework using a 3-fold cross-validation strategy where we train the model on two folds and test it on the left fold in an iterative manner. Since there is no framework aiming to learn an inter-domain alignment while superresolving brain graphs, we evaluated our proposed model L2S-KDnet with three ablated versions. All comparison methods are TS learning-based frameworks:
\begin{enumerate}
\item \textbf{Baseline:} we use a TS architecture where we do not include any adversarial regularization.
\item \textbf{Baseline+Discriminator:} a variant of the first method where we add a discriminator for distinguishing between the ground truth and predicted HR brain graphs.
\item \textbf{L2S-KDnet w/o Local Topology}: it is an ablated version of our L2S-KDnet framework which adopts only a global topology loss function.
\item \textbf{L2S-KDnet w/o TD regularization}: in this method, we remove the teacher's decoder from the training process of the student network.
\end{enumerate}

\begin{figure}[t!]
\centering
\includegraphics[width=12.5cm]{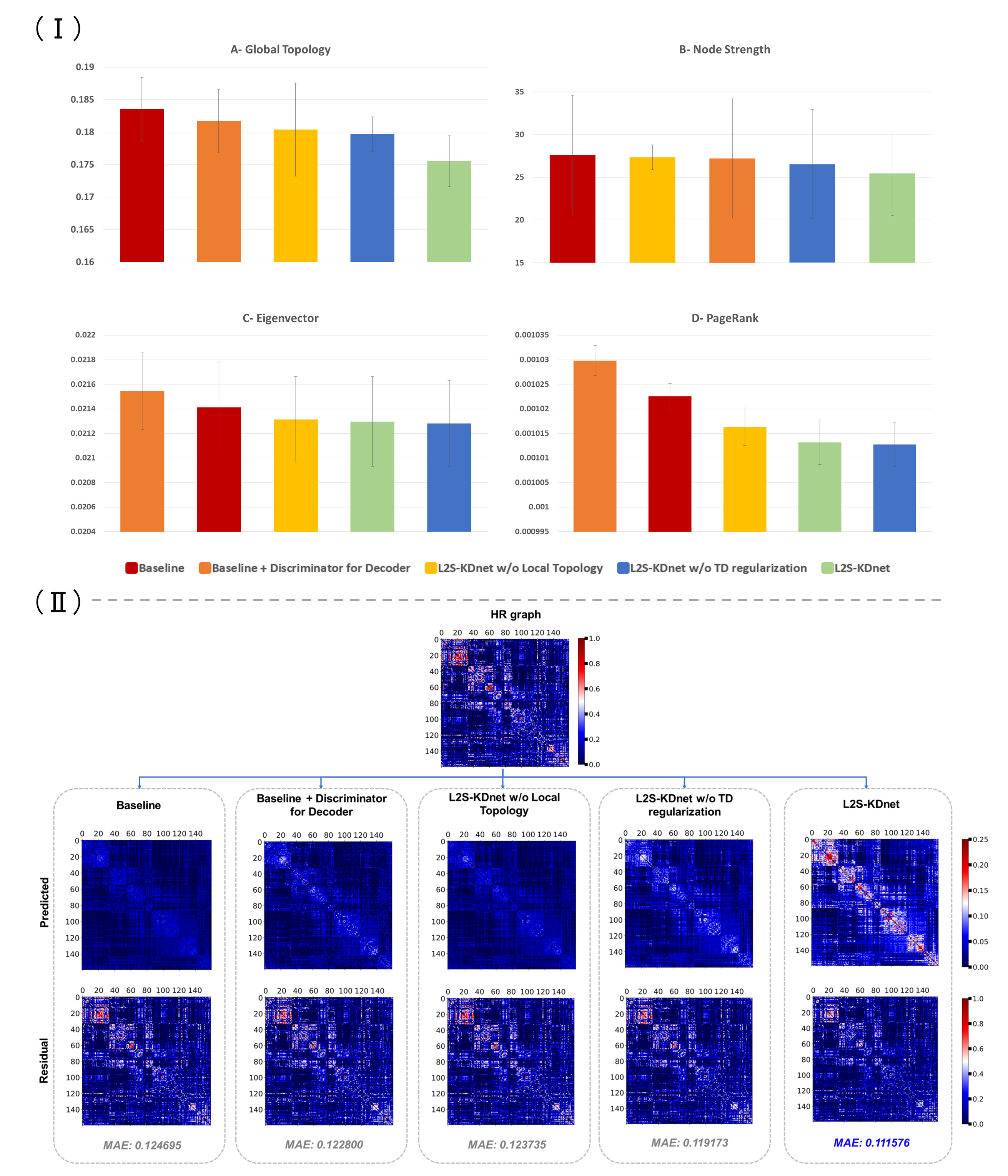}
\caption{ \emph{Comparison of student model of our L2S-KDnet framework against different ablated methods.} In \textbf{(\rom{1})} we plot the Mean Absolute Error (MAE) between ground truth and predicted HR brain graphs and the MAE of the node strength, eigenvector and PageRank centrality scores of the predicted and original graphs. In \textbf{(\rom{2})} we display the residual error computed using the mean absolute error (MAE) metric between the original HR brain graph and the predicted graph of a representative testing subject. 
}
\label{results_figure1}
\end{figure}

\textbf{Fig.}~\ref{results_figure1}\textbf{-(\rom{1})} shows the MAE between the ground truth and the predicted HR brain graphs and the MAE between the ground truth and synthesized centrality scores. In addition to the node strength measure, we further evaluate our framework using the eigenvector (EC) \cite{Bonacich:2007} and PageRank (PC) \cite{Brin:1998} centralities measures since they are commonly used for graph analysis \cite{Borgatti:2006}. We plot the average results over three folds of the student network as it is the model deployed in the testing stage. Our L2S-KDnet framework achieved the best prediction performance across all benchmark methods when using global topology and node strength evaluation measures (\textbf{Fig.}~\ref{results_figure1}-A,\textbf{Fig.}~\ref{results_figure1}-B). Interestingly, it outperformed the L2S-KDnet w/o Local Topology method which demonstrates the importance of our \emph{topology-aware adversarial loss} for training the teacher thereby distilling both global and local topological knowledge of the HR brain graphs to train the student network. 

While L2S-KDnet ranked first best when evaluated on global topology and node strength, it ranked second-best in MAE between the eigenvector and PageRank centralities as displayed in \textbf{Fig.}~\ref{results_figure1}-C,\textbf{Fig.}~\ref{results_figure1}-D. This shows that the incorporation of the teacher's decoder in the regularization of the student does not improve a lot the preservation of the local topological properties --in particular the eigenvector and PageRank centralities. This is explainable since the teacher's decoder acts as a generator in a GAN architecture, therefore it is exposed to the mode collapse issue \cite{Goodfellow:2014}. In other words, the decoder of the teacher might produce graphs that mimic a few modes of the original HR brain graphs, thus it solely captures a limited variability of the local topological properties. Designing a solution for circumventing the mode collapse issue of our teacher network is our future avenue. However, we argue that although the TD regularization method did not boost our model performance in preserving complementary local topological properties, our model still outperformed the three remaining competing methods and achieved the best performance when preserving the local topology depicted by the node strength measure. In \textbf{Fig.}~\ref{results_figure1}\textbf{-(\rom{2})}, we display the residual prediction error computed using MAE between the ground truth and predicted HR brain graphs for a representative subject. It shows that our framework yields a low residual in comparison to its ablated versions. More specifically, such results demonstrate the breadth of our inter-domain alignment-based teacher network which clearly boosts the network superresolution accuracy.

\section{Conclusion}

We proposed L2S-KDnet the first teacher-student framework designed for brain graph super-resolution based on inter-domain alignment learning. Our key contributions consist in: (i) designing a domain-aware teacher network based on a graph encoder-decoder able to predict a high-resolution brain graph from a low-resolution one, (ii) introducing a \emph{topology-aware adversarial loss function} using a node strength measure to enforce the teacher to preserve topological properties of the ground truth HR graphs, and (iii) proposing a novel way to train the student network which leverages the pretrained teacher decoder to boost the graph super-resolution result. Experiments on a healthy connectomic dataset showed that our proposed L2S-KDnet is able to achieve a low prediction error compared to its variants. In our future work, we plan to generalize our framework to a multiresolution brain graph synthesis task where given a single LR graph we can generate multiple HR graphs at different resolutions. Another interesting future direction is to leverage our framework to synthesize HR brain graphs for unhealthy subjects and then leverage the predicted graphs for boosting early disease diagnosis from low-resolution connectomes.

\section{Acknowledgements} 

This work was funded by generous grants from the European H2020 Marie Sklodowska-Curie action (grant no. 101003403, \url{http://basira-lab.com/normnets/}) to I.R. and the Scientific and Technological Research Council of Turkey to I.R. under the TUBITAK 2232 Fellowship for Outstanding Researchers (no. 118C288, \url{http://basira-lab.com/reprime/}). However, all scientific contributions made in this project are owned and approved solely by the authors. A.B is supported by the same TUBITAK 2232 Fellowship.

\section{Supplementary material}

We provide three supplementary items for reproducible and open science:

\begin{enumerate}
	\item A 5-mn YouTube video explaining how our framework works on BASIRA YouTube channel at \url{https://youtu.be/6RJebfo6ETc}.
	\item L2S-KDnet code in Python on GitHub at \url{https://github.com/basiralab/L2S-KDnet}. 
	\item A GitHub video code demo on BASIRA YouTube channel at \url{https://youtu.be/lvEfNG5AO_E}. 
\end{enumerate}

\bibliography{Biblio3}

\begin{thebibliography}{10}
\providecommand{\url}[1]{\texttt{#1}}
\providecommand{\urlprefix}{URL }
\providecommand{\doi}[1]{https://doi.org/#1}

\bibitem{Bassett:2017}
Bassett, D.S., Sporns, O.: Network neuroscience. Nature neuroscience
  \textbf{20}(3),  353--364 (2017)

\bibitem{GNNreviewAlaa}
Bessadok, A., Mahjoub, M.A., Rekik, I.: Graph neural networks in network
  neuroscience. arXiv preprint arXiv:2106.03535  (2021)

\bibitem{Bonacich:2007}
Bonacich, P.: Some unique properties of eigenvector centrality. Social networks
   \textbf{29}(4),  555--564 (2007)

\bibitem{Borgatti:2006}
Borgatti, S.P., Everett, M.G.: A graph-theoretic perspective on centrality.
  Social networks  \textbf{28}(4),  466--484 (2006)

\bibitem{Brin:1998}
Brin, S.: The {PageRank} citation ranking: bringing order to the web.
  Proceedings of ASIS, 1998  \textbf{98},  161--172 (1998)

\bibitem{kubra}
Cengiz, K., Rekik, I.: Predicting high-resolution brain networks using
  hierarchically embedded and aligned multi-resolution neighborhoods. In:
  Rekik, I., Adeli, E., Park, S.H. (eds.) Predictive Intelligence in Medicine.
  pp. 115--124. Springer International Publishing, Cham (2019)

\bibitem{chen2018efficient}
Chen, Y., Shi, F., Christodoulou, A.G., Xie, Y., Zhou, Z., Li, D.: Efficient
  and accurate {MRI} super-resolution using a generative adversarial network
  and 3d multi-level densely connected network. In: International Conference on
  Medical Image Computing and Computer-Assisted Intervention. pp. 91--99.
  Springer (2018)

\bibitem{Goodfellow:2014}
Goodfellow, I., et~al.: Generative adversarial nets. Advances in neural
  information processing systems pp. 2672--2680 (2014)

\bibitem{hinton2015distilling}
Hinton, G., Vinyals, O., Dean, J.: Distilling the knowledge in a neural
  network. arXiv preprint arXiv:1503.02531  (2015)

\bibitem{gsr}
Isallari, M., Rekik, I.: {GSR-Net}: Graph super-resolution network for
  predicting high-resolution from low-resolution functional brain connectomes.
  In: Liu, M., Yan, P., Lian, C., Cao, X. (eds.) Machine Learning in Medical
  Imaging. pp. 139--149. Springer International Publishing, Cham (2020)

\bibitem{megi}
Isallari, M., Rekik, I.: Brain graph super-resolution using adversarial graph
  neural network with application to functional brain connectivity. Medical
  Image Analysis  \textbf{71},  102084 (2021).
  \doi{https://doi.org/10.1016/j.media.2021.102084},
  \url{https://www.sciencedirect.com/science/article/pii/S1361841521001304}

\bibitem{Kipf:2016}
Kipf, T.N., Welling, M.: Semi-supervised classification with graph
  convolutional networks. arXiv preprint arXiv:1609.02907  (2016)

\bibitem{li2013automated}
Li, W., Andreasen, N.C., Nopoulos, P., Magnotta, V.A.: Automated parcellation
  of the brain surface generated from magnetic resonance images. Frontiers in
  neuroinformatics  \textbf{7}, ~23 (2013)

\bibitem{mihri}
Mhiri, I., Khalifa, A.B., Mahjoub, M.A., Rekik, I.: Brain graph
  super-resolution for boosting neurological disorder diagnosis using
  unsupervised multi-topology connectional brain template learning. Medical
  Image Analysis  \textbf{65},  101768 (2020).
  \doi{https://doi.org/10.1016/j.media.2020.101768},
  \url{https://www.sciencedirect.com/science/article/pii/S1361841520301328}

\bibitem{pilanci2020domain}
Pilanci, M., Vural, E.: Domain adaptation on graphs by learning aligned graph
  bases. IEEE Transactions on Knowledge and Data Engineering  (2020)

\bibitem{pilavci2019graph}
Pilavci, Y.Y., Guneyi, E.T., Cengiz, C., Vural, E.: Graph domain adaptation
  with localized graph signal representations. arXiv preprint arXiv:1911.02883
  (2019)

\bibitem{qiusouthwest}
Qiu, J., Qinglin, Z., Bi, T., Wu, G., Wei, D., Yang, W.: Southwest university
  longitudinal imaging multimodal (slim) brain data repository: A long-term
  test-retest sample of young healthy adults in southwest china

\bibitem{qu2020synthesized}
Qu, L., Zhang, Y., Wang, S., Yap, P.T., Shen, D.: Synthesized 7{T MRI from 3T
  MRI} via deep learning in spatial and wavelet domains. Medical image analysis
   \textbf{62},  101663 (2020)

\bibitem{sanchez2018brain}
S{\'a}nchez, I., Vilaplana, V.: Brain {MRI} super-resolution using {3D}
  generative adversarial networks. arXiv preprint arXiv:1812.11440  (2018)

\bibitem{shen2019brain}
Shen, Y., Gao, M.: Brain tumor segmentation on {MRI} with missing modalities.
  In: International Conference on Information Processing in Medical Imaging.
  pp. 417--428. Springer (2019)

\bibitem{soussia20197}
Soussia, M., Rekik, I.: 7 years of developing seed techniques for alzheimer’s
  disease diagnosis using brain image and connectivity data largely bypassed
  prediction for prognosis. In: International Workshop on PRedictive
  Intelligence In MEdicine. pp. 81--93. Springer (2019)

\bibitem{yang2021knowledge}
Yang, J., Martinez, B., Bulat, A., Tzimiropoulos, G.: Knowledge distillation
  via softmax regression representation learning. In: International Conference
  on Learning Representations (2021),
  \url{https://openreview.net/forum?id=ZzwDy_wiWv}

\bibitem{zhou2019limited}
Zhou, B., Lin, X., Eck, B.: Limited angle tomography reconstruction: Synthetic
  reconstruction via unsupervised sinogram adaptation. In: International
  conference on information processing in medical imaging. pp. 141--152.
  Springer (2019)

\end{thebibliography}
\bibliographystyle{splncs}
\end{document}